\documentclass{aa}
\usepackage{epsfig}

\def\power#1{\mbox{$\times10^{#1}\ $}}

\newcommand{\msun}{M$_\odot$ }
\newcommand{\msyr}{M$_\odot$.yr$^{-1}$ }

\newcommand{\zapj}{ApJ~}

\newcommand{\znp}{Nucl. Phys. }
\newcommand{\zprl}{Phys. Rev. Lett.}
\newcommand{\zpl}{Phys. Lett.}
\newcommand{\zpr}{Phys. Rev.}

\newcommand{\zmnras}{MNRAS}

\newcommand{\gap}{\mathrel{ \rlap{\raise.5ex\hbox{$>$}}
                      {\lower.5ex\hbox{$\sim$}}  } }
\newcommand{\lap}{\mathrel{ \rlap{\raise.5ex\hbox{$<$}}
		      {\lower.5ex\hbox{$\sim$}}  } }
\newcommand{\fo}{$^{18}$F}
\newcommand{\fn}{$^{19}$F}
\newcommand{\ox}{$^{17}$O}

\newcommand{\pg}{$^{18}$F(p,$\gamma)^{19}$Ne}
\newcommand{\pa}{$^{18}$F(p,$\alpha)^{15}$O}

\begin{document}

\thesaurus{06     % A&A Section 6: Form. struct. and evolut. of stars
              (02.14.1;  %  Nuclear reactions, nucleosynthesis, abundances
               08.14.2;  % novae, cataclysmic variables
               13.07.3)} % Gamma rays: theory
\title{Influence of new reaction rates on $^{18}$F production in novae}

\author{Alain Coc\inst{1} \and Margarita Hernanz\inst{2} 
\and Jordi Jos\'e\inst{2,3} \and Jean-Pierre Thibaud\inst{1}}
          
\institute{
Centre de Spectrom\'etrie Nucl\'eaire et de Spectrom\'etrie
de Masse, IN2P3-CNRS and Universit\'e Paris Sud, B\^atiment 104,\\ 
91405 Orsay Campus, France
\and
Institut d'Estudis Espacials de Catalunya/CSIC,
Edifici Nexus-201, C/Gran Capit\`a 2-4, E-08034 Barcelona, Spain
\and
Departament de F\'{\i}sica i Enginyeria Nuclear (UPC), Avinguda
V\'{\i}ctor Balaguer, s/n, E-08800 Vilanova i la Geltr\'u (Barcelona),
Spain
}

\date{Received 28 January 2000 / Accepted 9 March 2000}

\maketitle

\begin{abstract}
Gamma-ray emission from classical novae is dominated, during the first 
hours, by 
positron annihilation resulting from the beta decay of radioactive nuclei. 
The main contribution comes from the decay of \fo\ and hence is directly
related to \fo\ formation during the outburst. A good knowledge of the 
nuclear reaction rates of production and destruction of \fo\ is required
to study \fo\ synthesis in novae and the resulting gamma-ray emission. 
The rates relevant for the main mode of \fo\ destruction (i.e, through 
proton captures) have been the object of many
recent experiments. However, subsequent analyses were focused on 
providing rates for X-ray burst nucleosynthesis not valid at nova 
temperatures (lower than 3.5\power{8}~K).  
Accordingly, it is crucial to propose and discuss new 
reaction rates, incorporating all new experimental results, down to 
the domain of nova nucleosynthesis.
We show that in this temperature regime, the \pg\ and \pa\ 
reaction rates remain uncertain and deserve further 
experimental and theoretical efforts. 
Our hydrodynamic calculations including the new nuclear rates demonstrate 
that their impact
on \fo\ synthesis in nova explosions is quite large and, consequently, the
early gamma-ray emission from classical novae is also affected.


\end{abstract}

\section{Introduction}

Classical novae emit gamma-ray radiation at and below 511 keV during the 
early epochs after the explosion. 
This emission is produced by electron-positron annihilation in the expanding 
envelope, and the subsequent Comptonization of the resulting gamma-ray 
photons, and it shows a line, at 511 keV, and a continuum, between 20 and 
511 keV (\cite{Gom98}). 
The positrons responsible for this emission come mainly from the 
disintegration of \fo\ (\cite{Lei87}; \cite{Gom98}), because its lifetime 
($\tau$=158 min) is such that positrons are emitted at the ``right time",
i.e., when the expanding envelope starts to be transparent to gamma-ray 
radiation. 
Therefore, the amount of radiation emitted strongly depends on the \fo\ 
content of the nova envelope. 

The synthesis of \fo\ in novae depends largely on some key nuclear reaction
rates of \fo\ destruction and production which are far from being
well known. 
This is the case, in particular, of the \pg\ and \pa\ reactions. 
Recent experimental studies (\cite{Gra97}; \cite{Utk98})
drastically improved the knowledge of these reaction rates with respect to 
previous studies (\cite{WK82}).
In a recent paper (\cite{Her99}), we have analyzed the influence
of these rates (\cite{Utk98}) of \pg\ and \pa\ on the final yields
of \fo\ for different models of CO and ONe novae.
The effect of the new rates was important, since a factor of 10 reduction 
in the yields and in the resulting gamma-ray fluxes was obtained for all 
the models. 
Therefore, we concluded that a more detailed analysis of the reaction 
rates was necessary, in order to predict the gamma-ray emission of
classical novae. 

The rates proposed by Utku et al. (1998) were limited to  
relatively high resonance energies and temperature domains, more appropriate 
for temperatures typical of X--ray bursts than for those of classical 
novae.
The reason is that for their determination neither the influence of 
uncertainties on low energy 
resonance strengths (based on assumed reduced width) nor the effect of 
{\em all} low energy resonance tails were considered.
At higher temperatures the rates are more reliable since 
their main contributions come from two directly measured resonances. 
The purpose of this paper is to provide rates for the  
$^{18}$F(p,$\gamma)^{19}$Ne and $^{18}$F(p,$\alpha)^{15}$O reactions valid in
the domain of temperature of nova nucleosynthesis, incorporating the latest
experimental data. In addition to the nominal rate, we provide 
upper and lower limits. Since other nuclear reactions also affect 
\fo\ synthesis 
in novae, a global analysis is done, including results from the recent NACRE 
compilation (\cite{NACRE}). With the new rates affecting \fo\ synthesis, new 
nova models have been computed, in order to determine the mass of \fo\ they 
eject and the impact of the new yields on their early gamma-ray emission.
 
The organisation of this paper is the following. In section \ref{s:synt} we 
describe the nucleosynthesis of \fo. In section \ref{s:frates} we discuss 
in detail the \fo+p rates and the
corresponding uncertainties while in the following section (\ref{s:orates}),
we briefly discuss other recently published rates. In section \ref{s:yields}
we present new results that 
show the influence of these new rates on \fo\ production. Following the 
conclusion, an appendix gives some analytical approximations to the rates.

\section{ Synthesis of $^{18}$F in classical novae}
\label{s:synt}

In this Section, we will focus on the nuclear paths leading to $^{18}$F
synthesis, through a detailed analysis of a representative model of an
ONe nova : a 1.25 \msun white dwarf, accreting 
solar-like matter at a rate of $2 \times 10^{-10}$
\msyr, assuming a 50\% degree of mixing with the outermost core of
composition taken from Ritossa, Garc\'{\i}a--Berro \& Iben (1996) 
(mainly $^{16}$O and $^{20}$Ne with traces of other Ne, Na and Mg
isotopes).
The evolutionary sequences leading to a nova outburst have been computed 
with an updated version of the code SHIVA (see Jos\'e \& Hernanz 1998),
a one-dimensional, implicit, hydrodynamical code in Lagrangian formulation,
which follows the course of the explosion from the onset of accretion up 
to the expansion and ejection stages. The nuclear reaction network is 
described in Jos\'e et al. (1999), and in Hernanz et al. (1999). 
In particular, we use the reaction rates for proton captures on \fo\ and \ox\ 
based on Utku et al. (1998) and Landr\'e et al. (1989), respectively.

\begin{figure} % fig 1
\epsfig{file=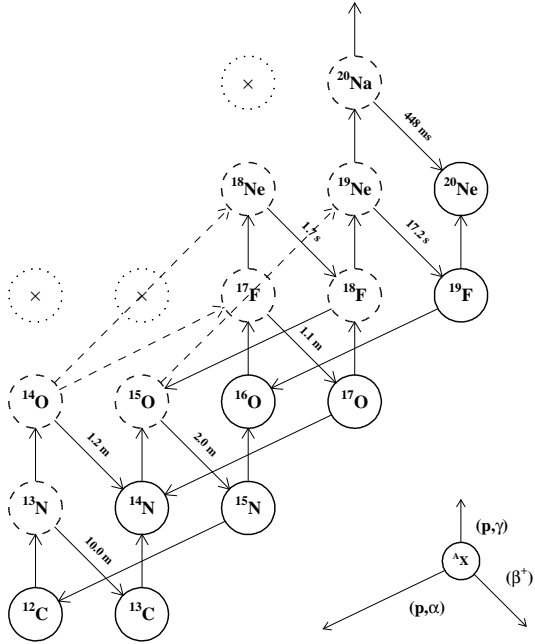,width=11.cm}
%\vspace{10pt}
\caption{Hot CNO cycle. 
Dashed and dotted circles represent beta unstable and proton unbound nuclei
respectively. Dashed arrows represent reactions of negligible influence 
in novae.}
\label{f:hcno}
\end{figure}

The synthesis of $^{18}$F in nova outbursts takes place within the hot-CNO
cycle (see fig. \ref{f:hcno}). Regardless of the nova type (either CO or ONe, 
according to the composition of the underlying white dwarf), the initial
$^{16}$O abundance is quite large (up to 25\% by mass). Hence, initial
$^{16}$O is the main source of the formation of $^{18}$F, either through
the chain $^{16}$O(p,$\gamma$)$^{17}$F(p,$\gamma$)$^{18}$Ne($\beta^+$)$^{18}$F
or through $^{16}$O(p,$\gamma$)$^{17}$F($\beta^+$)$^{17}$O(p,$\gamma$)$^{18}$F,
which reflects a competition between (p,$\gamma$) reactions and 
$\beta^+$-decays.

\begin{figure*} % fig 2
\begin{center}
\epsfig{file=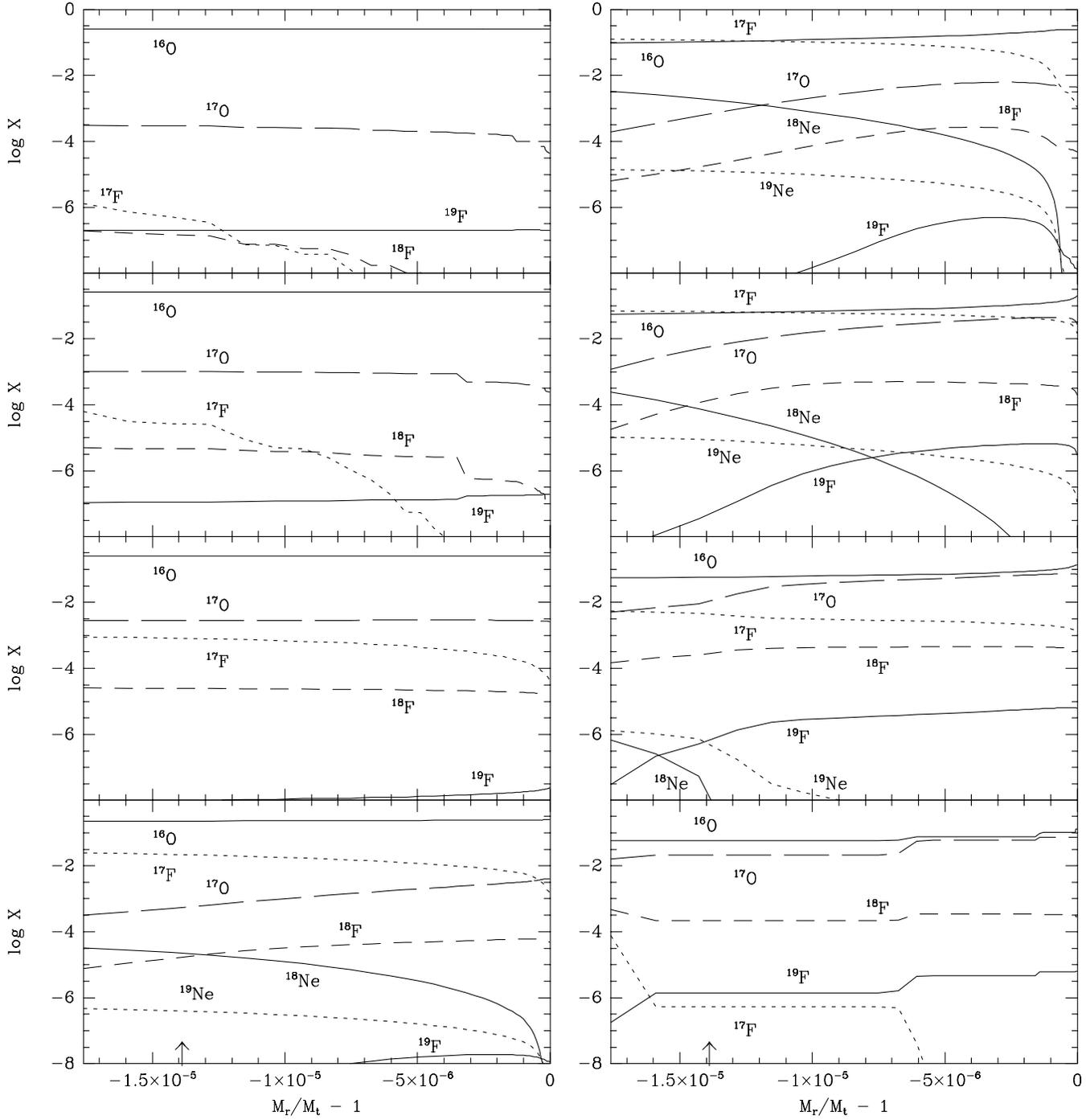,width=22.cm}
%\vspace{10pt}
\caption{Mass fractions of isotopes in the expanding envelope at various
stages of the outburst. The horizontal axis corresponds to a relative mass
coordinate where the origin is set at the surface of the envelope and
where the arrow shows the limit of the ejected material.
The panels are numbered from top to bottom and from left to right and
correspond to temperatures of approximately 5$\times10^7$~K,
7$\times10^7$~K, $10^8$~K, 2$\times10^8$~K, $T_{max}$=2.5$\times10^8$~K, 
and to the last 
phases of the evolution, when the nova envelope has already expanded to a size
of $R_{wd} \sim 10^9, 10^{10}$ and $10^{12}$ cm, respectively.  
}
\label{f:prof}
\end{center}
\end{figure*}

Snapshots of the evolution of isotopes relevant to $^{18}$F synthesis (i.e., 
$^{16,17}$O, $^{17,18,19}$F and $^{18,19}$Ne) are shown in Fig. \ref{f:prof}.
At the early stages of the explosion, when the temperature achieved at the 
burning shell reaches $T_{bs} \simeq 5 \times 10^7$ K (Fig. \ref{f:prof}, 
first panel), the main nuclear activity in this mass region is driven by 
$^{16}$O(p,$\gamma$)$^{17}$F, followed by its $\beta^+$-decay to \ox. 
However, the temperature is not high enough to burn a significant fraction 
of $^{16}$O. On the contrary, the abundance of \ox\ rises by several 
orders of magnitude with respect to its initial content. 
Both $^{17,18}$F are being synthesized at the burning shell by means of 
proton captures on $^{16,17}$O, respectively, whereas the amount of $^{19}$F 
remains essentially unchanged. The minor activity driven by proton capture 
reactions onto $^{17,18}$F results also in a moderate increase of both 
$^{18,19}$Ne (below $10^{-10}$ by mass).

A similar trend is found when $T_{bs} \simeq 7 \times 10^7$ K (Fig. 
\ref{f:prof}, second panel). 
The temperature rise increases the number of proton captures
onto $^{16}$O, leading to $^{17}$F, which rapidly decays into \ox.
The effect of convection, which already extends through most of the envelope
(i.e., $\sim 150$ km), is shown in the smooth distribution of both $^{17,18}$F, 
previously synthesized in the burning shell and efficiently carried away 
to the outer envelope. $^{19}$F is in turn reduced down to 50\% due to 
$^{19}$F(p,$\alpha$)$^{16}$O, which dominates $^{19}$Ne($\beta^+$)$^{19}$F. 

When $T_{bs}$ reaches $10^8$ K (Fig. \ref{f:prof}, third panel), \ox\ 
shows a flat profile along the envelope, with a mean mass fraction of 
$\sim 3 \times 10^{-3}$, due to 
$^{16}$O(p,$\gamma$)$^{17}$F($\beta^+$)$^{17}$O, which dominates destruction
through $^{17}$O(p,$\gamma$)$^{18}$F. With respect to fluorine isotopes, we 
stress an important synthesis of $^{17,18}$F at this stage, driven by 
$^{16}$O(p,$\gamma$)$^{17}$F and $^{17}$O(p,$\gamma$)$^{18}$F, whereas 
$^{19}$F is almost fully destroyed. $^{18,19}$Ne continue to rise due to
proton captures onto $^{17,18}$F.

A dramatic change is found as soon as the burning shell reaches 
$2 \times 10^8$~K (Fig. \ref{f:prof}, fourth panel). At this time, the mass
fraction of $^{16}$O has been reduced down to $\sim 3 \times 10^{-2}$. For the 
first time, \ox\ decreases (by one order of magnitude with respect to 
the amount shown in previous panel), since destruction through 
$^{17}$O(p,$\gamma$)$^{18}$F dominates $^{17}$F($\beta^+$)$^{17}$O in the 
vicinity of the burning shell. On the contrary, $^{17}$F exhibits a 
significant rise, up to $\sim 3 \times 10^{-2}$ by mass (i.e., one order 
of magnitude higher than in previous panel), so that the amount of $^{17}$F
becomes larger than that of \ox\ in almost all the envelope. Again, 
$^{16}$O(p,$\gamma$)$^{17}$F dominates both $^{17}$O(p,$\gamma$)$^{18}$Ne and 
$^{17}$F($\beta^+$)$^{17}$O  near the burning shell. The evolution of $^{19}$F
is still dominated by  $^{19}$F(p,$\alpha$)$^{16}$O, which in fact has no 
significant influence on the $^{16}$O content. Also noticeable is the 
dramatic rise of both $^{18,19}$Ne, which increase by several orders of 
magnitude, mainly due to the fact that at such temperatures, proton captures
onto $^{17,18}$F become faster than the corresponding $\beta^+$-decays.
It is worth noticing that at this stage, a non-negligible path leading 
to $^{18}$F  synthesis is now driven by $^{18}$Ne($\beta^+$)$^{18}$F, 
basically at the outer envelope shells, which compensates an efficient 
destruction through proton captures near the burning shell.  

The burning shell achieves a maximum temperature of $2.51 \times 10^8$ K 
(Fig. \ref{f:prof}, fifth panel). $^{16}$O is reduced to 0.096 by mass in 
the burning shell. Whereas \ox\ is destroyed by (p,$\alpha$) and 
(p,$\gamma$) reactions at the burning shell, its mean abundance in the 
envelope increases due to $^{17}$F($\beta^+$)$^{17}$O which dominates 
in almost all the envelope.
On the contrary, $^{17}$F is generated in the burning shell by means of 
$^{16}$O(p,$\gamma$)$^{17}$F, whereas it is being destroyed by $\beta^+$-decays 
in the outer envelope. The evolution of $^{18}$F reflects a competition between
destruction and creation modes at different locations of the envelope: whereas
its content is reduced by both $^{18}$F(p,$\alpha$)$^{15}$O reactions at the 
burning shell and by $^{18}$F($\beta^+$)$^{18}$O at the outer envelope, 
a dominant source for $^{18}$F synthesis through $^{17}$O(p,$\gamma$)$^{18}$F 
is found at the intermediate shells. 
Proton captures onto $^{17,18}$F and convective transport continue to pump 
$^{18,19}$Ne to the outer envelope shells, which is at the origin 
of the rise of $^{19}$F at this stage (through $^{19}$Ne($\beta^+$)$^{19}$F).

Shortly after, due to the sudden release of energy from the short-lived species
$^{13}$N, $^{14,15}$O and $^{17}$F, the envelope begins to expand. As a result
of the drop in temperature, (p,$\gamma$) and (p,$\alpha$) reactions are
basically restricted to the vicinity of the burning shell, whereas most of the
envelope is dominated by $\beta^+$-decays (Fig. \ref{f:prof}, sixth and 
seventh panels).
Hence, whereas \ox\ is powered by $^{17}$F($\beta^+$)$^{17}$O along the 
envelope, the amount of $^{17}$F and $^{18,19}$Ne decreases significantly.
The $^{18}$F abundance at these late-time stages of the outburst increases 
due to $^{17}$O(p,$\gamma$)$^{18}$F, which is dominant at the intermediate 
layers of the envelope (whereas in the vicinity of the burning shell $^{18}$F 
is efficiently destroyed by $^{18}$F(p,$\alpha$)$^{15}$O, the outer envelope 
is dominated by $^{18}$F($\beta^+$)$^{18}$O). Moreover, destruction of 
$^{19}$F through $^{19}$F(p,$\alpha$)$^{16}$O at the burning shell is 
dominated by $^{19}$Ne($\beta^+$)$^{19}$F elsewhere. 

The final stages of the outburst (Fig. \ref{f:prof}, eighth panel), as the 
envelope expands and cools down, are dominated by the release of nuclear 
energy by $\beta^+$-decays such as $^{18}$F($\beta^+$)$^{18}$O, or 
$^{17}$F($\beta^+$)$^{17}$O. The resulting mean abundance of $^{18}$F in 
the ejected shells in this Model is X($^{18}$F)= $2.7 \times 10^{-4}$.
Most of the envelope is, however, dominated by the large abundances of 
$^{16,17}$O (X($^{16}$O)= $6.8 \times 10^{-2}$, X($^{17}$O)= 
$3.9 \times 10^{-2}$). A residual $^{17}$F, which is still decaying into
$^{17}$O, and a non-negligible amount of $^{19}$F ($3 \times 10^{-6}$, by 
mass) are also present.

In summary, the synthesis of $^{18}$F in classical novae 
is essentially controlled by five proton-capture reactions, 
$^{16}$O(p,$\gamma$)$^{17}$F, 
$^{18}$F(p,$\gamma$)$^{19}$Ne, $^{18}$F(p,$\alpha$)$^{15}$O, 
$^{17}$O(p,$\gamma$)$^{18}$F, $^{17}$O(p,$\alpha$)$^{14}$N, 
and several $\beta^+$-decays. 
Accordingly, the corresponding reaction rates deserve further attention
and their uncertainties are discussed in Sect.~\ref{s:frates} for the \fo+p
rates and Sect.~\ref{s:orates} for the other rates.

\section{The \pa\ and \pg\ reaction rates}
\label{s:frates}

Uncertainties on the \pg\ and \pa\ reaction rates used to reach many
orders of magnitude due to the very limited spectroscopic data available for 
$^{19}$Ne in the domain of interest. 
The rate estimated by Wiescher \& Kettner (1982) has become obsolete by
the recent measurements of Rehm et al. (1995), Coszach et al. (1995),
Rehm et al. (1996, 1997)
%\cite{Reh95,Cos95,Reh96,Reh97} 
and in particular by the work of Graulich et al. (1997), Utku et al. (1998)
and Butt et al. (1998). 
%\cite{Gra97,Utk98,But98}. 
We concentrate on the faster
$^{18}$F(p,$\alpha)^{15}$O reaction but most of the discussion also applies
to the $^{18}$F(p,$\gamma)^{19}$Ne one.

\begin{figure} % fig 3
\epsfig{file=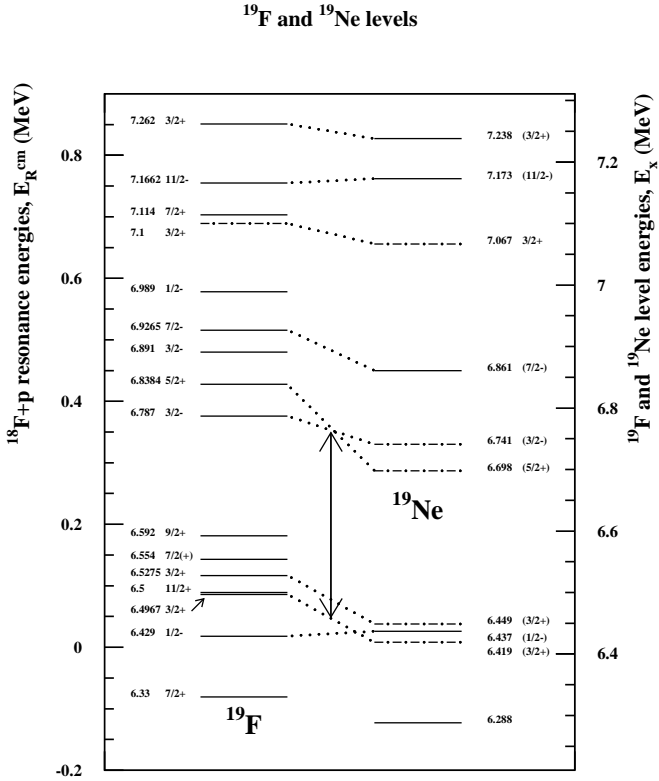,width=9.cm}
%\vspace{10pt}
\caption{Level scheme for \fn\ and $^{19}$Ne. Dotted lines relate
conjugate levels suggested by Utku et al. 1998. 
The arrow represents the energy range where the presence of resonances (or
tails of resonances) affects the reaction rates in the domain of temperature
corresponding to nova outbursts. 
The dash-dotted lines represents new experimental data since Wiescher \&
Kettner (1982).}
\label{f:levels}
\end{figure}

Before 1997 and below the excitation energy ($E_x$) corresponding to the 
resonance energy ($E_r$) of 0.4~MeV, only two levels were known 
$E_x$ = 6.437 and 6.742~MeV but with unknown or uncertain spins and parity 
($J^\pi$)\footnote{For more detailed explanations of the standard nuclear 
physics notations used, see e.g. the appendix in Jos\'e et al. (1999).}.
On the contrary, various levels were known in the corresponding region
of \fn, the conjugate nucleus (see Fig.~\ref{f:levels}).
For instance, in the $E_r$=0.--1.~MeV region, only 6 levels were known and
many spins unknown in $^{19}$Ne, while 19 levels had been observed in the
corresponding \fn\ region.
Hence, for the estimation of the rates, Wiescher \& Kettner (1982) 
considered three 
known levels at $E_x$ = 6.437, 6.742 and 6.862~MeV and postulated three 
others.
The corresponding rate was highly uncertain since the location of these 
three postulated levels was approximate and because other levels found in
\fn\ are expected to have counterparts in $^{19}$Ne in the region of
interest. 
This is the case, in particular, of the $E_x,J^\pi$ =  6.429, 1/2$^-$, 
6.497, 3/2$^+$ and 6.527, 3/2$^+$ levels of \fn\ corresponding to strong
expected  resonances because of their low centrifugal barrier, i.e. low
transfered orbital angular momentum ($\ell_p$) of 1, 0 and 0 respectively.

Even though the recent experiments have not been able to find all the
counterparts of the \fn\ levels (see Fig.~\ref{f:levels}),
they provided direct measurement for the
strengths of two resonances which likely dominate the rate in the domain
of nova nucleosynthesis and the location of several new $^{19}$Ne levels.
The strength of the resonance corresponding to the $E_x$ = 6.742~MeV level
($E_r$=330~keV)
has recently been measured directly by Graulich et al. (1997) with a \fo\ 
beam provided by the Louvain--La--Neuve facility.
This level is thought to be the analog of the 6.787, 3/2$^-$ \fn\ level.
The measured strength ($\omega\gamma$),
$3.5\pm1.6$~eV (\cite{Gra97}), is in good agreement with the Wiescher \&
Kettner (1982) estimate and accordingly  does not induce a significant 
change in the rate.
On the contrary, the resonance associated with the  $E_x,J^\pi$ =
7.067~MeV, 3/2$^+$ level ($E_r$=659~keV) strongly modifies the rates.
It has been studied by 
Rehm et al. (1995, 1996), Coszach et al. (1995),
Graulich et al. (1997), Utku et al. (1998) and Butt et al. (1998) 
%\cite{Reh95,Reh96,Cos95,Gra97,Utk98,But98} 
and its strength has been measured directly (\cite{Cos95}; \cite{Gra97}).
It is located well outside of the Gamow peak but due to its total width
($\Gamma\approx$30~keV), the contribution of its tail alone is greater
than the Wiescher \& Kettner (1982) rate in the domain of nova nucleosynthesis.
However, no counterpart of this broad level was known in \fn.
The closest 3/2$^+$ \fn\ level is located at $E_x$ = 7.262 but its width
is much smaller ($\Gamma<$6~keV).
To remove this incompatibility, Rehm et al. (1996) claimed that the 
7.067~MeV width could be smaller ($\Gamma\approx$14~keV).
A 7.114, 7/2$^+$ level was known in \fn\ with the possibility that
it hides an unresolved $J^\pi$ = 3/2$^+$ level as suggested by 
Smotrich et al. (1961).
A recent gamma ray spectroscopy study (\cite{But98}) of \fn\ produced by
the $^{15}$N($\alpha,\gamma)^{19}$F reaction has confirmed the
presence of a 7.101~MeV, 3/2$^+$, $\Gamma$ = $28\pm1$~keV level.
Hence the high width found by Coszach et al. (1995) and Utku et al. (1998) 
%\cite{Cos95,Utk98} 
for the $E_x,J^\pi$ =
7.067~MeV, 3/2$^+$,  $^{19}$Ne level can be understood. (Note however that 
this analog assignment has been questioned very recently by 
Fortune \& Sherr (2000).)

\begin{figure*} % fig 4
\begin{center}
\epsfig{file=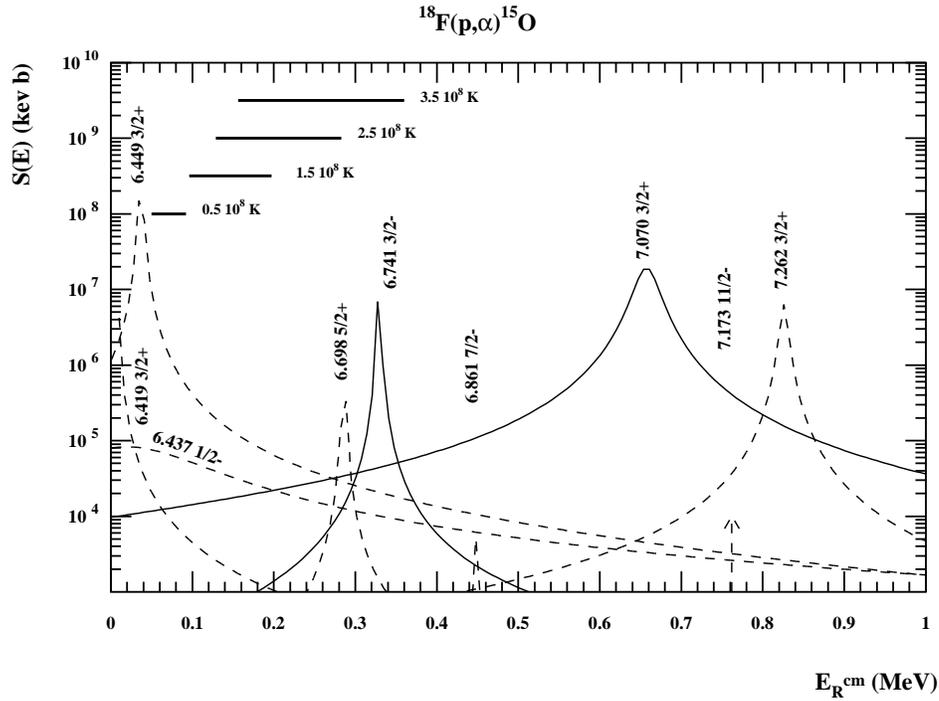,width=13.cm}
%\vspace{10pt}
\caption{Astrophysical S-factor for the $^{18}$F(p,$\alpha)^{15}$O
reaction. The solid lines represent contributions from resonances
whose strengths have been directly measured. 
Dashed curves represent assumed 
contributions from resonances corresponding to known
$^{19}$Ne levels.
Possible contributions associated to expected levels in $^{19}$Ne are
not shown. Note the importance of the tail of the $E_r$=38 keV 
($E_x$=6.449 MeV, 3/2$^+$) resonance not considered in Utku et al. (1998).}
\label{f:pase}
\end{center}
\end{figure*}

Utku et al. (1998) have also found three new $^{19}$Ne levels and have made 
tentative assignments of analog levels in \fn\ (see Fig.\ref{f:levels}).
They recognized the 6.437~MeV, 1/2$^-$ level ($E_r$=26~keV) as the analog 
of the 6.429, 1/2$^-$, $\Gamma$ = 280~keV one in \fn\ from its large measured 
width ($216\pm19$~keV). 
Even though it is located at a very low energy (in the context of novae)
it is so broad that its tail
may lead to a significant contribution to the rate (see Fig.\ref{f:pase}). 
The 6.449~MeV 3/2$^+$ ($E_r$=38~keV) level is one of the three new ones 
found by Utku et al. (1998). 
It is also of great importance since the contribution of its tail
can dominate the astrophysical factor 
in the relevant energy range (see Fig.\ref{f:pase}).
In comparison the two other new levels give a smaller
(6.698~MeV 5/2$^+$, $E_r$=287~keV) or negligible contribution 
(6.419~MeV 3/2$^+$, $E_r$=8~keV).

\begin{figure*}[h] % fig 5
\begin{center}
\epsfig{file=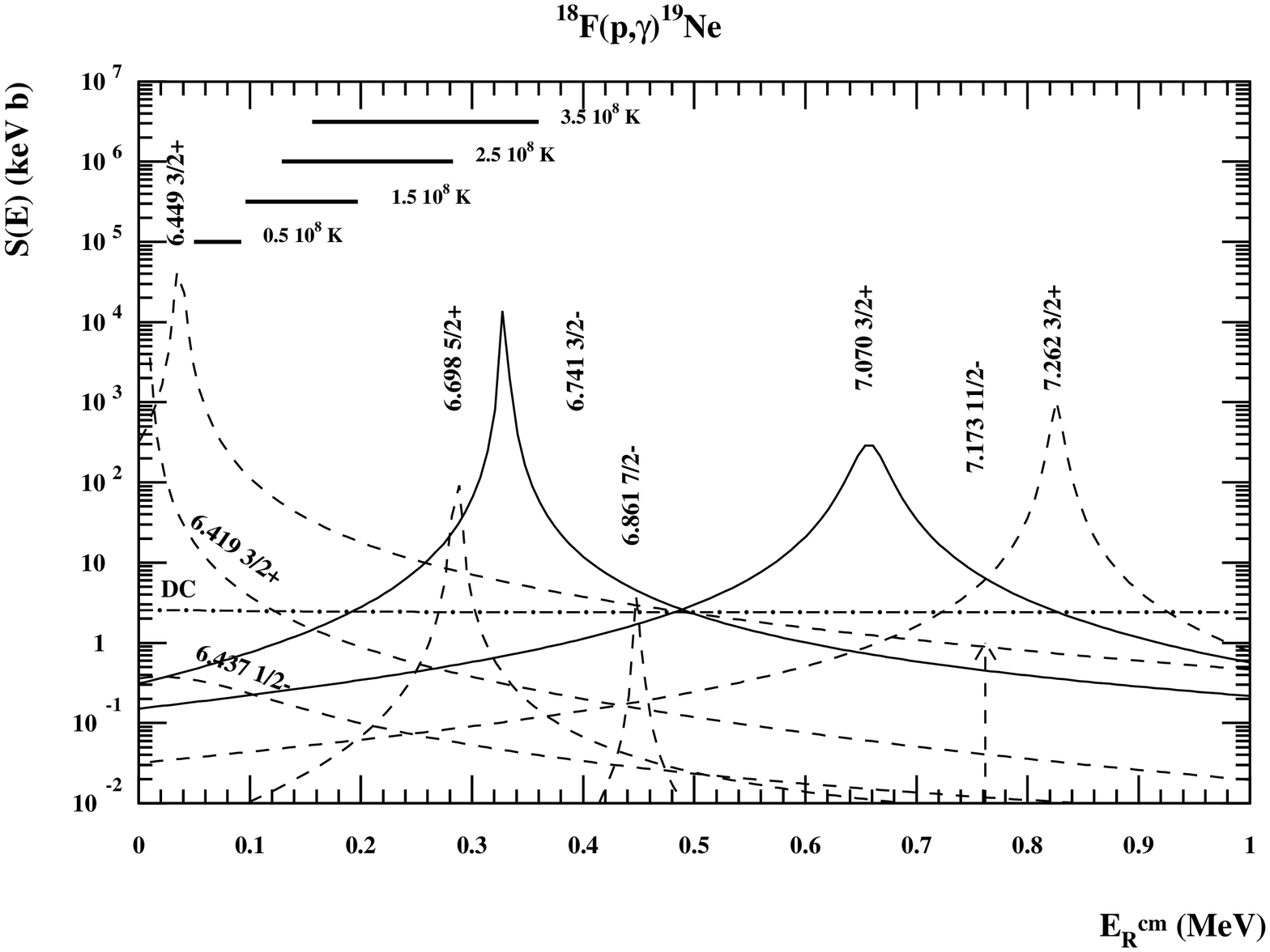,width=13.cm}
%\vspace{10pt}
\caption{Same as Fig.~\protect\ref{f:pase} but for the
$^{18}$F(p,$\gamma)^{19}$Ne 
reaction.}
\label{f:pgse}
\end{center}
\end{figure*}

A new rate has been provided by Utku et al. (1998), using their new data 
together with 
available spectroscopic information on $^{19}$Ne and \fn\ supplemented
by estimates of radiative and proton widths when missing.
This rate is already higher by a factor of up to $\approx3\times10^2$ when
compared with Wiescher \& Kettner (1982). 
Nevertheless, even though this rate is now set on a firmer basis than the
Wiescher \& Kettner (1982) one, it remains quite uncertain as it depends
directly on the estimated proton widths.
Except for the two resonances whose strengths have been measured directly
by Graulich et al. (1997), 
the proton widths are assumed to be equal to 0.1 or 0.01
Wigner limit, i.e. $\theta^2_p$=0.1 or 0.01, according to their parity 
(\cite{Utk98}). 
For the resonances considered here, one has
$\Gamma_p\ll\Gamma_{Tot}\approx\Gamma_\alpha$ so that their strengths are
($\omega\gamma\approx\omega\Gamma_p$) proportional to chosen $\theta^2_p$ 
values.
Accordingly, to obtain lower and upper limits for the rate, we used the two
extreme cases $\theta^2_p$=0 and $\theta^2_p$=1 respectively and
regardless of parity but kept Utku et al. (1998) $\theta^2_p$ values for 
the nominal rate (see Fig.\ref{f:parate}).
The lower limit is quite certain as it is given by the contribution of the
two directly measured resonances whose parameters are now reliably determined.
It remains higher than the Wiescher \& Kettner (1982) rate above 
$\approx$50\power{6}~K.
The nominal rate is conditioned by the rather arbitrary choice of
$\theta^2_p$=0.1 or 0.01 that we adopt following Utku et al. (1998).
However, our nominal rate is significantly higher [up to a factor of
$\approx$10 and 30 around 10$^8$~K for (p,$\alpha$) and (p,$\gamma$) 
reactions]
than the one from Utku et al. (1998) for typical nova temperatures
(see Fig.\ref{f:parate}). This is due to the inclusion of the contribution of
the tail ($\Gamma_{Tot}=\Gamma_\alpha$=4.3~keV) of the 6.449~MeV level 
(see Fig.~\ref{f:pase}) not considered by Utku et al. (1998).
The corresponding rates (p,$\alpha$) are presented in Fig.\ref{f:parate}
and have been calculated by numerical integration of the Breit-Wigner
formula for all broad resonances.
For the (p,$\gamma$) rates, we used the same procedure to obtain the low,
nominal and high rates (see Figs.~\ref{f:pgse} and \ref{f:pgrate}).
Radiative widths were taken from \fn\
analog levels as in Utku et al. (1998) except for the important resonance at 659~keV
as Bardayan et al. (1999) provided an experimental value 
($\Gamma_\gamma$=0.39~eV) for the analog level. 
We also adopted the direct capture (DC) contribution from Utku et al. (1998)
even though the spectroscopic factors used ($^{18}$O$\otimes$p instead of 
$^{18}$F$\otimes$p or even $^{18}$F$\otimes$n) 
may not be the more appropriates.  
These rates have been calculated using all the available spectroscopic data
collected in Table~II of Utku et al. (1998) except for the inclusion of a new gamma
width for the 659~keV resonance and for a different proton width for the 
26~keV resonance [1\% of the Wigner limits i.e. 2.5$\times10^{-17}$ eV, 
as for other negative parity resonances, instead of 6.6$\times10^{-20}$ eV 
of unknown origin in Utku et al. (1998)]. 
These numbers are taken at face value to calculate the nominal rate only. For the low and 
high rates unknown proton widths are allowed to vary between 0 and 
the Wigner limit as discussed above. 
It is clear that this range ($0<\theta^2_p<1$) represents the most extreme 
values as are the corresponding rates. This should be kept in mind when
interpreting the astrophysical implications.

However  a few points have been neglected due to the lack of experimental
or theoretical information.
First of all, the identification of all $E_x$ $\lap$ 6.6~MeV analog levels
is not complete yet (see Fig.~\ref{f:levels}) and is not certain. In
particular, new contributions to the nominal and high rates corresponding to
missing levels cannot be ruled out.
The tail of the 38~keV resonance has been calculated using the large total 
width derived from the \fn\ analog level proposed by Utku et al. (1998). 
Another 3/2$^+$ level, with smaller width, lies 30~keV below in 
\fn. It has been assigned (\cite{Utk98}) to be the analog level
corresponding to the 8~keV (E$_x$=6.419~MeV) resonance in \fo+p 
(see Fig.~\ref{f:levels}). It is possible that the analog assignment for 
these two levels is inversed or that they are mixed but at least one or 
the other resonance (8 or 38~keV) should be broad enough to dominate the rates
at low nova temperature.
For the (p,$\gamma$) rates, we neglected any difference of radiative widths
between analog levels and the uncertainty on the DC contribution in front of
the uncertainty on the proton widths.
These rates do not either include possible interference effects
arising from the interference between the 3/2$^+$ resonances
located at 38 and 659~keV. In a favourable case
(when $\theta^2_p\approx$0.01 for the 38~keV resonance), destructive
interference could reduce the low rate by a factor of $\approx$10 in the 
sensitive $T\approx10^8$~K region.
However, at present the rates provided in the appendix are those we recommend.

\section{Other reactions affecting \fo\ production}
\label{s:orates}

According to the recent compilation of charged-particle induced
thermonuclear reaction rates NACRE (Angulo et al. 1999),
the rate for $^{16}$O(p,$\gamma$)$^{17}$F is rather well known and 
suffers from a little uncertainty (i.e., a factor of $\sim$ 2 between the
high and low rates). 
Even though $^{17}$F has a much shorter lifetime than \fo, its
destruction by proton capture is rather well known at nova temperatures and
should not affect the analysis of Sect.~\ref{s:synt}.
Below $\approx$4\power{8}~K, the $^{17}$F(p,$\gamma)^{18}$Ne reaction rate is 
dominated by the effect of direct capture on bound $^{18}$Ne levels. 
This contribution is expected to suffer little uncertainty since it has been 
calculated 
(\cite{WK82}; \cite{Gar91}) using experimental neutron spectroscopic factors 
from the mirror nucleus $^{18}$O.
The lowest resonances are located at $E_r$ = 0.586, 0.655 and 0.600~MeV 
(\cite{Bar99}) too high to contribute, even by their tails, 
to the rate below $\approx$4\power{8}~K. 
Partial widths are reasonably known through experiments,
shell model calculations or from the conjugate levels (\cite{WK82}; 
\cite{Gar91}) and are such that $\Gamma_p\gg\Gamma_\gamma$
(i.e. $\omega\gamma=\omega\Gamma_\gamma$).
A long standing uncertainty (\cite{WK82}; \cite{WGT88}; \cite{Gar91}) 
concerned the 
location of the $\ell$=0, 0.600~MeV resonance associated with the 
4.524~MeV, 3$^+$ level. 
Fortunately, the resonance energy first measured by Garc\'{\i}a et al. (1991), 
has been experimentally confirmed very recently by Bardayan et al. (1999).
Hence the uncertainty associated with the $^{17}$F(p,$\gamma)^{18}$Ne rate
appears negligible within the domain of nova nucleosynthesis. 

\begin{figure} 
\begin{center}
\epsfig{file=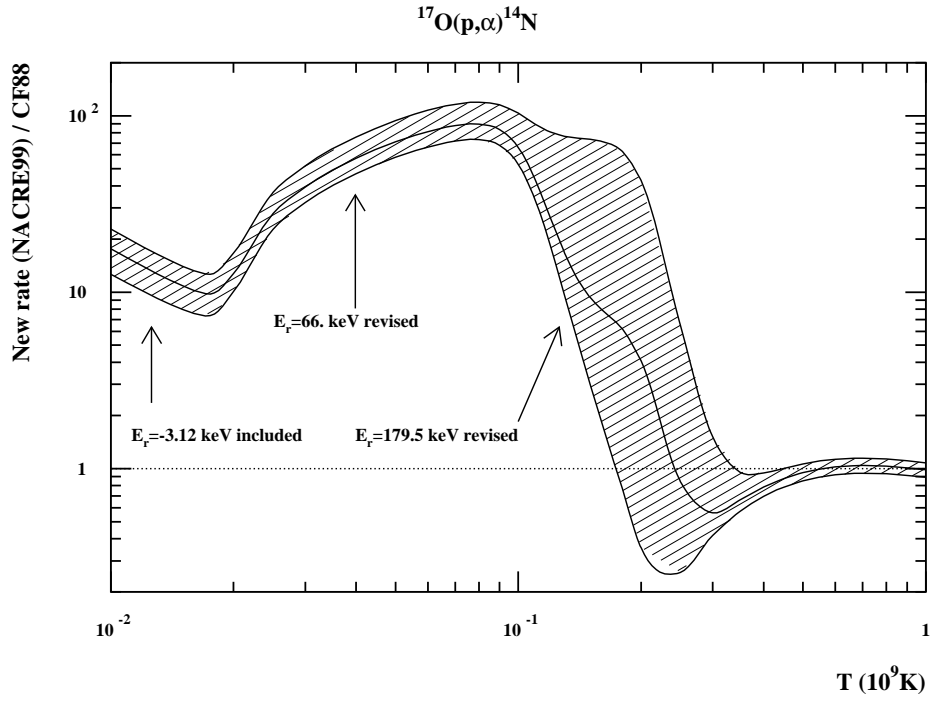,width=13.cm}
%\vspace{10pt}
\caption{Comparison between NACRE (low, recommended and high) and CF88 rates 
for the $^{17}$O(p,$\alpha)^{14}$N reaction. The origins for the
differences are labeled with the energy of the corresponding resonances.
(See \protect\cite{NACRE} for more detail.)}
\label{f:o17pa}
\end{center}
\end{figure}

\begin{figure}
\begin{center}
\epsfig{file=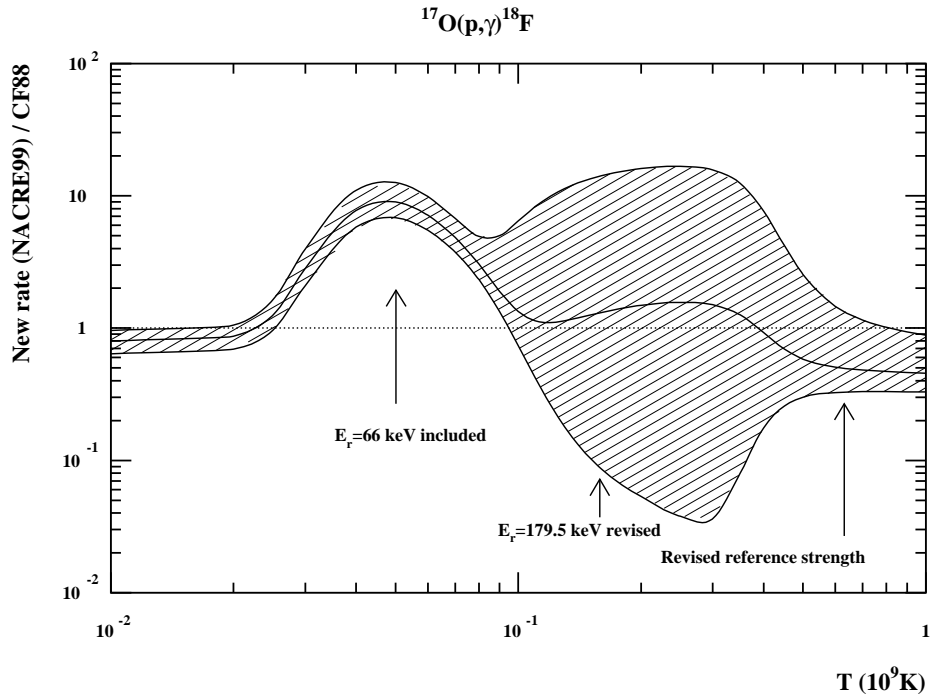,width=13.cm}
%\vspace{10pt}
\caption{
Same as Fig.~\protect\ref{f:o17pa} but for the
$^{17}$O(p,$\gamma)^{18}$F 
reaction.}
\label{f:o17pg}
\end{center}
\end{figure}

According to NACRE (\cite{NACRE}), the $^{17}$O(p,$\alpha)^{14}$N
(Fig.~\ref{f:o17pa})
and $^{17}$O(p,$\gamma)^{18}$F
(Fig.~\ref{f:o17pg})
reaction rates present large
uncertainties at temperatures below a few 10$^8$~K.
The increase in the rates around 50\power{6} (\cite{NACRE}) is due to the
contribution of a $E_r$ = 66~keV resonance (\cite{Lan89}; \cite{Bla95}),
not included in Caughlan \& Fowler (1988).
The large uncertainty around 2\power{8}~K (\cite{NACRE}), well within the 
range of temperature reached in nova outbursts, is due to the unknown 
contribution of a 179.5~keV resonance associated with the 
E$_{x}$ = 5786 keV ($^{18}$F) level. 
The upper limits for the strengths [(p,$\gamma$) and (p,$\alpha$)] of this 
resonance are calculated (\cite{NACRE}) from the partial widths
extracted by Rolfs, Berka \& Azuma (1973) and the upper limit for the 
proton width from Landr\'e et al. (1989).
As usual, for the calculation of the contribution of this                
resonance to the NACRE low, recommended and high rates, its strength has 
been taken equal to 0, 10\% and 100\% of the experimental upper limit.
Accordingly, the NACRE recommended rate is somewhat arbitrary in the
temperature domain affected by the 179.5~keV resonance (Figs.~\ref{f:o17pa}
and Figs.~\ref{f:o17pg})

\section{Results: effect of new rates on \fo\ synthesis}
\label{s:yields}

New models of a 1.25 \msun ONe nova have been computed with the SHIVA code, in 
order to analyze the effect of the new nuclear reaction rates. For the \fo+p 
rates, we have adopted the low, high and nominal prescriptions described in 
section 3, whereas for the \ox+p ones we have adopted the NACRE rates (see  
Table~\ref{t:yields}). It is important to stress that the energetics of the 
explosion remains practically unchanged with the new \fo+p and \ox+p reaction 
rates, since these reactions are not the ones affecting most the evolution 
(see Jos\'e \& Hernanz (1998) and Jos\'e et al. (1999) for a deep analysis
of the reaction rates which have the largest influence in nova explosions). 
Therefore, the maximum 
temperature attained in the burning shell at the base of the accreted envelope, 
T$_{max}$=2.51\power{8}~K, the mean kinetic energy and the ejected mass, 
$\Delta$$M_{ejec}$=1.8\power{-5}\msun, are unchanged with respect to previous 
models (i.e., Jos\'e et al. (1999)). 

Two previously computed evolutionary sequences, with the same white 
dwarf mass and composition, are adopted as {\it old} models for the purpose 
of comparison: a case with Wiescher \& Kettner (1982) rates 
(see Jos\'e \& Hernanz (1998) for the yields and G\'omez-Gomar et al. (1998)
for the $\gamma$-ray spectra) and a case with Utku et al. (1998) rates 
(see Hernanz et al. (1999) for the yields and the spectra). 
In both of them, the Caughlan \& Fowler (1988) 
updated by Landr\'e et al. (1989) rates for \ox+p were adopted. 
In Table~\ref{t:yields} we show the \fo\ yields obtained with the new rates, as 
well as those with the {\it old} models. We see that yields with the nominal 
\fo+p rates are smaller by a factor of 30 than those with the older rates, 
as expected. In fact, the reduction in \fo\ production is even larger (by a 
factor of 
3) than the one obtained with Utku et al. (1998) rates (see Hernanz et al.
(1999)), as a consequence of the inclusion of the tail of the resonance at
38 keV (see section 3).

Concerning the effect of the new \ox+p rates, a reduction by an extra factor 
of 2 is obtained when the NACRE recommended rates, instead of the old
Caughlan \& Fowler (1988) and Landr\'e et al. (1989) ones, 
are adopted (together with \fo+p nominal rates; see Table~\ref{t:yields}). 
In summary, our nominal 
%best number for the 
yield of \fo\ is 4.84\power{-5}, 
which is 60 times smaller than the yield with the old Wiescher \& Kettner 
(1982) rates (\cite{Jos98}; \cite{Gom98}) 
and 6 times smaller than that with Utku et al. (1998) rates (\cite{Her99}). 
The consequences of the reduced \fo\ yields are quite large for the early 
gamma-ray output, which 
is directly related to the amount of \fo\ synthesized. The reduced content of 
\fo\ in an expanding envelope with similar physical and dynamical properties, 
translates directly into a reduced positron annihilation gamma-ray flux, 
roughly by the same factor as the \fo\ decrease. This was shown in 
Hernanz et al. (1999), 
where complete computations of the gamma-ray spectra were performed with the 
\fo\ yields obtained with Utku et al. (1998) rates, for different masses and 
compositions of the white dwarf. With the rates presented here, the reduction 
of the predicted fluxes becomes larger by a factor of 6, for the combined 
nominal \fo+p and recommended \ox+p rates. In summary, if we compare with 
results in G\'omez-Gomar et al. (1998), where \fo\ yields were computed 
with Wiescher \& Kettner (1982), 
the 511~keV and continuum fluxes for a 1.25~\msun\ ONe nova should be reduced 
by a factor of 60.

Another important point to stress is the impact of the uncertainty in the
rates on the computed yields. The above {\em nominal} yields
rely upon {\em recommended} or {\em nominal} rates which contain assumed
values ($\theta^2_p$=0.01 or 0.1 for \fo+p and 10\% of an experimental
upper limit for \ox+p). Hence, the nominal yield should not be dissociated
from the large remaining uncertainties.  
From our results (see Table~\ref{t:yields}) we conclude 
that the range between high and low presciptions for the \fo+p rates translates 
into an uncertainty (maximum versus minimum yield ratio) of a factor of 310, 
a quite large value, whereas for the \ox+p rates the uncertainty is a factor 
of 10. (It is worth noticing that the results obtained with the 
Utku et al. (1998) rates 
fall within the limits obtained with the new rates; see Table~\ref{t:yields}).
The corresponding uncertainties in the gamma-ray fluxes are of the same 
order of magnitude. This points out the interest of more accurate
determinations of the \fo+p and \ox+p rates.

\section{Conclusions}

We have investigated the \fo\ formation and destruction in nova          
outbursts, identified the key reactions (proton capture on \fo\ and \ox)
and analysed their rates.
The proton capture rates on \fo\ are higher than the Wiescher \&
Kettner (1982) ones at nova
temperatures due to the contribution of the tail of the 659 keV resonance
whose large measured width (\cite{Cos95}; \cite{Gra97}) has been
indirectly confirmed (\cite{But98}).
Another important contribution comes from the tail
of the 38~keV resonance which was neglected in previous studies.
Its {\em nominal} contribution is larger than the 659 keV one but
is proportional to its {\em assumed} reduced proton width $\theta^2_p$.
The strengths (proton widths) of the low lying resonances are unknown 
and induce large uncertainties (factors of 100 to 1000) in the rates 
at nova temperatures. We have provided updated 
nominal rates for the two capture reactions together with upper and lower 
limits. 
The \ox+p rates also display some large uncertainties at nova temperatures
according to the recent compilation of Angulo et al. (1999).

We have used these new nuclear physics results in a fully hydrodynamical nova 
code to calculate the \fo\ yields in novae for different rates : our low, 
nominal and high \fo+p rates, and the low, recommended and high \ox+p rates 
from Angulo et al. (1999). 
These results have been compared with models computed with the 
old Wiescher \& Kettner (1982) rates (\cite{Jos98}), and with more 
recent models with the Utku et al. (1998) rates (\cite{Her99}). Two 
important results have been obtained. First, there is always a reduction of the 
amount of \fo\ synthesized in a nova explosion, with the nominal rates for 
the \fo+p reactions both alone and combined with the recommended 
rates for the \ox+p reactions. The nominal \fo\ yield
%final \fo\ yield for this best assumption 
(nominal \fo+p and recommended for \ox+p) is 4.84\power{-5} by mass, which is 
60 times 
smaller than the one obtained with Wiescher \& Kettner (1982) rates 
and 6 times smaller than the one with Utku et al. (1998) rates 
(and Caughlan \& Fowler (1988) and Landr\'e et al. (1989) for \ox+p). 
The impact on the early gamma-ray spectrum of the nova is a reduction of the 
flux by the same amount (with respect to G\'omez-Gomar et al. (1998) and 
Hernanz et al. (1999), respectively). 
Second, the yields are found to be very sensitive to the rates 
with resulting combined (\fo+p and \ox+p) uncertainties of more than three 
orders of magnitude. This supports the need of new experimental and theoretical 
studies to improve the knowledge of the \fo+p and \ox+p rates and, 
consequently, allow for a larger reliability of the predictions of 
annihilation gamma--ray fluxes from novae, to be observed by current and 
future instruments.

\begin{acknowledgements}
This work was partially supported by PICS 319, PB98-1183-C02,            
PB98-1183-C03 and ESP98-1348.
\end{acknowledgements}

\newpage
\onecolumn

\appendix
%\subsection{Appendix: Updated rates}
{\bf Appendix: Updated rates}

\begin{figure}[h] % fig 6
\begin{center}
\epsfig{file=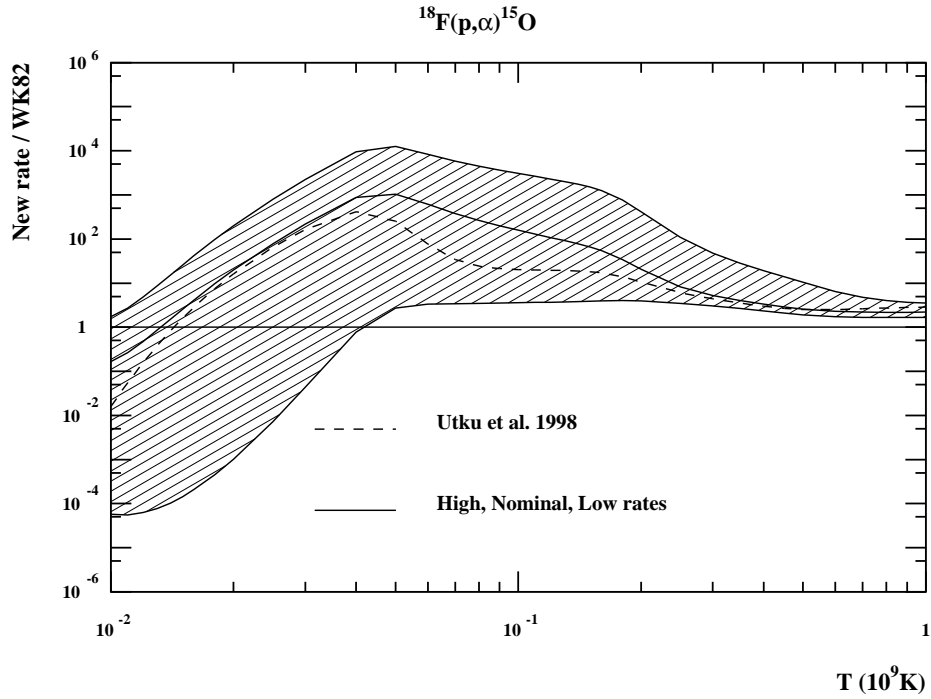,width=13.cm}
%\vspace{10pt}
\caption{$^{18}$F(p,$\alpha)^{15}$O 
rates compared to WK82. The solid lines represent the low, nominal
and high rates and the hatched area the nuclear uncertainty. The dashed line
represents the 
\protect\cite{Utk98} 
rate. All rates have been normalized
to the WK82 one.}
\label{f:parate}
\end{center}
\end{figure}

\begin{figure*}[h] % fig 7
\begin{center}
\epsfig{file=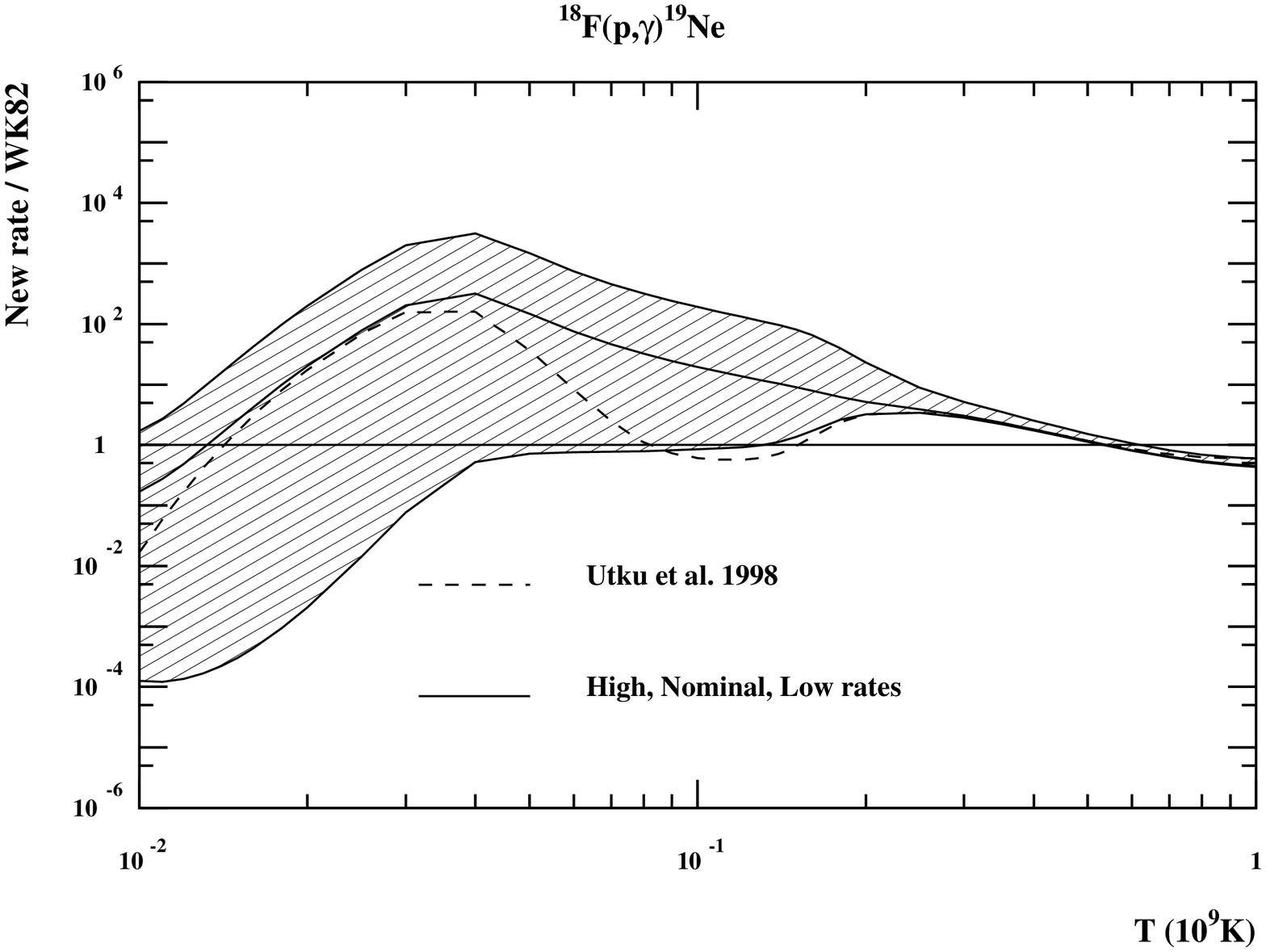,width=13.cm}
%\vspace{10pt}
\caption{Same as Fig.~\protect\ref{f:parate} but for the
$^{18}$F(p,$\gamma)^{19}$Ne 
reaction.}
\label{f:pgrate}
\end{center}
\end{figure*}

We give here the tabulated reaction rates for proton capture on \fo\ 
resulting from numerical integrations (see \S~\ref{s:frates} and 
Figs.~\ref{f:parate} and \ref{f:pgrate}).
For convenience, we also provide here formulas that approximate the nominal 
rates (As usual, T9 is the temperature in GK,
T9XY=T9**(X/Y) and T9LN=LOG(T9) in FORTRAN notations.) 

The following \pa\ nominal rate formula includes the contribution of the 
two measured resonances (330 and 659 keV) with additional contributions 
from the possible resonances at 38~keV, (with high energy tail) and 
287~keV.

\begin{verbatim}
      SVDir = 9.13e10/T923*EXP(-18.052/T913-0.672*T92)*(1.+! Tail
     & 0.0231*T913+6.12*T923+0.988*T9+9.92*T943+4.07*T953) ! 659 keV
     & +5.78e5/T932*EXP(-3.830/T9)+9.91e8/T932*EXP(-7.648/T9)+ ! 330, 659 keV
     & 2.81e-6/T932*EXP(-0.441/T9)+2706.*EXP(3.8319*T9LN-1.3450* ! 38 keV 
     & T9LN**2-.0001/T9**3)+4.46e4/T932*EXP(-3.331/T9) ! 287 keV
\end{verbatim}

The following \pg\ nominal rate formula contains contributions from direct
capture and from the 38, 287, 330 and 659~keV resonances.

\begin{verbatim}
      SVDir = 3.98e7/T923*EXP(-18.052/T913)* ! DC
     & (1.+0.0231*T913+0.0885*T923+0.0143*T9)+
     & 1.34e3/T932*EXP(-3.830/T9)+1.51e4/T932*EXP(-7.648/T9)
     & +EXP(-0.56781+3.6850*T9LN-1.3636*T9LN**2-.0001/T9**3)
     & +8.26e-10/T932*EXP(-0.441/T9)+12.2/T932*EXP(-3.331/T9)
\end{verbatim}

\newpage

\begin{table*}[ht]
\begin{center}
\caption{Yields (in mass fraction) of \fo\, 1 hour after T$_{max}$, for a 
1.25 \msun ONe nova, for different 
prescriptions for the \fo+p and \ox+p nuclear reaction rates.}
\begin{tabular}{lccc} 
\hline\hline\
Rates for \fo+p & Rates for \ox+p  & Reference
                & \fo\ \\
\hline
WK82            & CF88+Lan89       & JH98   
                & $2.80\times 10^{-3}$ \\
Utku98          & CF88+Lan89       & HJC99
                & $2.50\times 10^{-4}$ \\
Low             & CF88+Lan89       & This work   
                & $8.19\times 10^{-4}$ \\
Nominal         & CF88+Lan89       & This work    
                & $9.78\times 10^{-5}$ \\
High            & CF88+Lan89       & This work    
                & $2.98\times 10^{-6}$ \\
Nominal         & NACRE low        & This work    
                & $1.58\times 10^{-5}$ \\
Nominal         & NACRE rec.       & This work    
                & $4.84\times 10^{-5}$ \\
Nominal         & NACRE high       & This work 
                & $1.67\times 10^{-4}$ \\
\hline
\label{t:yields} 
\end{tabular}
\newline
References for the rates: WK82 (\cite{WK82}), CF88 (\cite{CF88}), 
Lan89 (\cite{Lan89}), NACRE (\cite{NACRE}). References for the models: JH98 
(\cite{Jos98}), HJC99 (\cite{Her99}).
\end{center}
\end{table*} 

\begin{table*}[h]
\caption[]{\pg\ and \pa\ rates}
\begin{tabular}{|r|r|r|r|r|r|r|}
\hline\noalign{\smallskip}
&\multicolumn{3}{c|}{\pa}&\multicolumn{3}{c|}{\pg}\\
\hline\noalign{\smallskip}
T (10$^9$~K) & low & nominal & high & low & nominal & high  \\
\hline\noalign{\smallskip}
 0.030& 0.323E-16& 0.841E-13& 0.841E-12& 0.605E-13& 0.302E-09& 0.308E-08\\
 0.040& 0.555E-14& 0.338E-11& 0.338E-10& 0.108E-10& 0.123E-07& 0.132E-06\\
 0.050& 0.215E-12& 0.445E-10& 0.444E-09& 0.431E-09& 0.165E-06& 0.197E-05\\
 0.060& 0.352E-11& 0.349E-09& 0.348E-08& 0.723E-08& 0.133E-05& 0.180E-04\\
 0.070& 0.328E-10& 0.195E-08& 0.194E-07& 0.689E-07& 0.765E-05& 0.116E-03\\
 0.080& 0.208E-09& 0.843E-08& 0.835E-07& 0.444E-06& 0.339E-04& 0.568E-03\\
 0.090& 0.991E-09& 0.296E-07& 0.292E-06& 0.216E-05& 0.122E-03& 0.221E-02\\
 0.100& 0.383E-08& 0.883E-07& 0.865E-06& 0.843E-05& 0.372E-03& 0.717E-02\\
 0.110& 0.126E-07& 0.231E-06& 0.225E-05& 0.279E-04& 0.990E-03& 0.201E-01\\
 0.120& 0.367E-07& 0.544E-06& 0.523E-05& 0.810E-04& 0.237E-02& 0.502E-01\\
 0.130& 0.992E-07& 0.118E-05& 0.112E-04& 0.212E-03& 0.519E-02& 0.114E+00\\
 0.140& 0.262E-06& 0.239E-05& 0.222E-04& 0.516E-03& 0.106E-01& 0.238E+00\\
 0.150& 0.703E-06& 0.467E-05& 0.416E-04& 0.120E-02& 0.203E-01& 0.464E+00\\
 0.160& 0.194E-05& 0.899E-05& 0.748E-04& 0.271E-02& 0.372E-01& 0.857E+00\\
 0.180& 0.145E-04& 0.344E-04& 0.222E-03& 0.135E-01& 0.114E+00& 0.255E+01\\
 0.200& 0.867E-04& 0.138E-03& 0.618E-03& 0.622E-01& 0.323E+00& 0.657E+01\\
 0.250& 0.251E-02& 0.291E-02& 0.656E-02& 0.143E+01& 0.339E+01& 0.457E+02\\
 0.300& 0.237E-01& 0.256E-01& 0.436E-01& 0.129E+02& 0.224E+02& 0.205E+03\\
 0.350& 0.115E+00& 0.121E+00& 0.182E+00& 0.630E+02& 0.957E+02& 0.669E+03\\
 0.400& 0.366E+00& 0.383E+00& 0.537E+00& 0.212E+03& 0.302E+03& 0.173E+04\\
 0.450& 0.887E+00& 0.921E+00& 0.124E+01& 0.577E+03& 0.794E+03& 0.382E+04\\
 0.500& 0.178E+01& 0.184E+01& 0.242E+01& 0.140E+04& 0.189E+04& 0.758E+04\\
 0.600& 0.489E+01& 0.504E+01& 0.648E+01& 0.662E+04& 0.879E+04& 0.248E+05\\
 0.700& 0.990E+01& 0.102E+02& 0.130E+02& 0.240E+05& 0.317E+05& 0.689E+05\\
 0.800& 0.167E+02& 0.172E+02& 0.220E+02& 0.675E+05& 0.887E+05& 0.165E+06\\
 0.900& 0.251E+02& 0.258E+02& 0.337E+02& 0.153E+06& 0.201E+06& 0.341E+06\\
 1.000& 0.349E+02& 0.361E+02& 0.481E+02& 0.295E+06& 0.387E+06& 0.623E+06\\
 1.250& 0.649E+02& 0.679E+02& 0.982E+02& 0.939E+06& 0.123E+07& 0.189E+07\\
 1.500& 0.101E+03& 0.107E+03& 0.169E+03& 0.196E+07& 0.258E+07& 0.394E+07\\
 1.750& 0.140E+03& 0.151E+03& 0.259E+03& 0.321E+07& 0.424E+07& 0.655E+07\\
 2.000& 0.180E+03& 0.197E+03& 0.361E+03& 0.454E+07& 0.601E+07& 0.948E+07\\
 2.500& 0.262E+03& 0.292E+03& 0.581E+03& 0.703E+07& 0.936E+07& 0.155E+08\\
 3.000& 0.347E+03& 0.390E+03& 0.800E+03& 0.897E+07& 0.120E+08& 0.209E+08\\
\hline\noalign{\smallskip}
\end{tabular}
\label{t:rates} 
\end{table*}
\noindent

\end{document}